\documentclass[a4paper,twoside]{article}

\usepackage{calc}
\usepackage{multicol}
\usepackage{pslatex}
\usepackage{apalike}
\usepackage{epsfig}
\usepackage{algorithm2e}
\usepackage[bottom]{footmisc}
\usepackage{dirtytalk}
\usepackage{enumitem}
\usepackage{array}
\usepackage{listings}
\usepackage{framed}
\usepackage{xcolor}
\usepackage{booktabs}
\usepackage{graphicx}
\usepackage{SCITEPRESS}     

\begin{document}

\title{Beyond Static Scoring: Enhancing Assessment Validity via AI-Generated Interactive Verification}

\author{\authorname{Tom Lee\sup{1}}
\authorname{Sihoon Lee\sup{2}}
\authorname{Seonghun Kim\sup{3}\thanks{Corresponding author}}
\affiliation{\sup{1}Chongshin University, Industry-Academic Cooperation Foundation, Seoul, Republic of Korea}
\affiliation{\sup{2}Cheongju National University of Education, Cheongju, Republic of Korea}
\affiliation{\sup{3}Chongshin University, College of Liberal Arts Education, Seoul, Republic of Korea}
\email{toomy0toons@gmail.com, shoon1984@gmail.com, ryankim@csu.ac.kr}
}


\abstract{Large Language Models (LLMs) challenge the validity of traditional open-ended assessments by blurring the lines of authorship. While recent research has focused on the accuracy of automated scoring (AES), these static approaches fail to capture "process evidence" or verify genuine student understanding. This paper introduces a novel Human-AI Collaboration framework that enhances assessment integrity by combining rubric-based automated scoring with AI-generated, targeted follow-up questions. In a pilot study with university instructors ($N=9$), we demonstrate that while Stage 1 (Auto-Scoring) ensures procedural fairness and consistency, Stage 2 (Interactive Verification) is essential for construct validity—effectively diagnosing superficial reasoning or unverified AI use. We report on the system's design, instructor perceptions of "fairness vs. validity," and the necessity of adaptive difficulty in follow-up questioning. The findings offer a scalable pathway for authentic assessment that moves beyond policing AI to integrating it as a synergistic partner in the evaluation process.}
\onecolumn \maketitle \normalsize \setcounter{footnote}{0} \vfill

\section{\uppercase{Introduction}}
\label{sec:introduction}

The recent proliferation of Large Language Models (LLMs) such as ChatGPT has introduced significant challenges to academic integrity and assessment authenticity \cite{morjaria2023examining}. Students can now use AI to generate fluent, human-like text for open-ended assignments, making it difficult for educators to distinguish original work from AI-generated content \cite{morjaria2023examining}. This ambiguity undermines assessment fairness, potentially giving some students an unfair advantage while subjecting others to undue suspicion. In educational measurement, fairness is a multi-dimensional concept encompassing distributive, procedural, and interactional justice \cite{rasooli2025students}. Furthermore, assessment validity is known to be compromised by factors like response bias and design flaws \cite{quansah2024validity}. We contend that the ambiguity of authorship in the LLM era threatens both fairness and validity, risking a shift where assessments measure not \say{what a student knows}, but rather \say{what AI can produce}.

Existing countermeasures have significant limitations. Traditional Automated Essay Scoring (AES) systems often focus on surface-level linguistic features, failing to capture a student's true comprehension \cite{yavuz2025utilizing}. Consequently, a well-written but poorly understood submission can receive a high score \cite{yavuz2025utilizing}. Simply prompting an LLM to grade is also insufficient; without structured rubrics and few-shot examples, out-of-the-box models struggle with accuracy compared to human graders \cite{piscitelli2025large}. AI detection tools have proven to be \say{neither accurate nor reliable}, with high false positive and false negative rates that misclassify student work \cite{weber2023testing}. Moreover, recent research demonstrates that simple prompt engineering techniques can easily bypass these detectors \cite{ernst2025detect}, rendering static detection futile. While alternative methods like oral exams can verify authenticity, they are difficult to scale in large courses and impose a significant practical burden on both instructors and students \cite{fenton2024reconsidering}. This creates a clear need for a solution that is scalable, fair, and valid.

To address this gap, we propose a two-stage, instructor-in-the-loop assessment framework that synthesizes the efficiency of automation with the rigor of interactive verification. The process unfolds in two stages:

\begin{itemize}[nosep]
  \item{\textbf{Stage 1: Rubric-Based Automated Scoring.} The system first generates an analytic rubric from course materials and uses it to perform an initial, automated evaluation of a student's submission.}
  \item{\textbf{Stage 2: Interactive Verification and Reassessment.} Based on the initial scoring, the system generates targeted follow-up questions designed to probe the student's reasoning. Functioning as a scalable \say{oral defense}, this stage allows for a reassessment that incorporates the student's responses. If an initial submission is AI-generated or based on superficial understanding, its limitations are likely to become apparent during this second stage.}
\end{itemize}

This approach operationalizes the pedagogical practice of verifying understanding through dialogue \cite{morjaria2023examining}. It aims to disincentivize over-reliance on AI and strengthen validity by grounding scores in verifiable evidence of a student's thought process.

The main contributions of this paper are as follows:

\begin{itemize}[nosep]
  \item{\textbf{An Automated Two-Stage Framework}} that integrates automated scoring with an interactive verification loop to improve assessment fairness and validity.
  \item{\textbf{A Structured Method for Evidence-Based Verification}} that systematizes the ad hoc process of questioning students into a scalable, transparent workflow.
  \item{\textbf{Empirical Insights from an Instructor Pilot Study}} that provide preliminary evidence of the framework's perceived utility, fairness, and potential for classroom implementation.
\end{itemize}

This paper details our system's design, reports findings from the instructor pilot, and discusses the implications for creating more trustworthy assessments in the LLM era.

\section{\uppercase{Related Work}}

\subsection{Validity and Fairness in Pedagogy}

Validity and fairness are foundational criteria in educational assessment. Validity concerns whether an assessment accurately measures what a student knows and can do \cite{messick1995validity}, while fairness ensures that all students have an equal opportunity to demonstrate their learning \cite{glaser1997assessment}. These are not merely psychometric properties; they are deeply intertwined with the pedagogical design of learning experiences \cite{boud2007rethinking}, serving as cornerstones for both educational credibility and learner rights.

The rise of digital learning and LLMs has introduced new threats to these values. Researchers argue that while automated assessment can improve scoring reliability, it must be paired with human expertise and explainability to ensure valid and fair evaluation of student achievement \cite{burstein2024assessment}. Others note that AI-generated responses can obscure a student's actual understanding, thus harming validity, and that assessments must incorporate "process evidence" to be authentic \cite{chai2024grading}.

Therefore, the \say{automated scoring + follow-up questions} approach proposed in this paper is not merely an attempt to improve technical accuracy but a bid to simultaneously secure two pedagogically important goals: validity and fairness. Rubric-based scoring enhances procedural fairness by applying the same criteria, while follow-up questions supplement achievement validity by securing evidence of the learner's understanding and reasoning processes. This is not just an improvement in efficiency but a distinctive contribution linked to the fundamental values of educational assessment.

In addition, when interpreted through Messick's (1995) validity framework \cite{messick1995validity}, the two-stage procedure offers evidence across multiple dimensions. Stage 1 (rubric-based automated scoring) contributes to content validity and supports internal structure, while Stage 2 (interactive verification and reassessment) provides response process evidence and addresses consequential as well as generalizability aspects. Together, these stages also reinforce the social consequences of fairness and trust in evaluation. This mapping shows that the proposed framework not only responds to current challenges but also provides concrete pathways for gathering authenticity evidence in practice. At the same time, this interpretation is consistent with Glaser et al.'s (2001) perspective on fairness \cite{glaser1997assessment} as providing all learners with equal opportunity to demonstrate achievement, suggesting that the framework simultaneously addresses both validity and fairness at their theoretical foundations.

\subsection{Automated Essay Scoring (AES) and AI Detectors}

Automated Essay Scoring (AES) systems were developed to increase the efficiency and consistency of grading written work. Early systems like e-rater and Coh-Metrix focused on linguistic and syntactic features such as grammar, vocabulary, and organization to train machine learning models that emulate human raters \cite{deane2013relation,amorim2018automated}. More recent approaches leverage pre-trained language models like BERT and Transformers to better capture semantic meaning and argumentation structure \cite{nie2025automated,chassab2024optimized}, with some research focusing on improving performance in cross-prompt environments \cite{xu2025epcts}.

Despite these advances, AES systems have known limitations. They can overemphasize surface-level features at the expense of evaluating critical thinking \cite{deane2013relation}, exhibit bias against students from certain linguistic backgrounds \cite{schaller2024fairness}, and often lack generalizability across different assignment types.

In response to the rise of LLMs, AI text detection tools like GPTZero, Turnitin's AI detector, DetectGPT \cite{mitchell2023detectgpt}, and GLTR \cite{gehrmann2019gltr} have emerged. However, studies have shown their performance to be unreliable, with frequent false positives and false negatives that depend on text length and topic \cite{liang2023}. Furthermore, their effectiveness is easily compromised by simple evasion techniques like paraphrasing or translation, making them unsuitable as a standalone solution for ensuring academic integrity \cite{ippolito2019automatic}.

\subsection{Multimodal Approaches and Alternative Attempts}

To overcome the limitations of AES and AI detectors, some have proposed alternative methods like oral interviews or multimodal data analysis (e.g., voice, eye-tracking) to directly verify a student's knowledge \cite{lindblom2006self}. Deep learning-based multimodal analysis, in particular, shows promise for accurately capturing cognitive response patterns to verify authenticity \cite{choudhary2025beyond}. However, these approaches are often resource-intensive, difficult to scale, and raise significant privacy and ethical concerns related to biometric data collection \cite{jin2024fate}.

A more scalable alternative has emerged from research on automated follow-up questions. This approach is pedagogically valuable, as well-designed questions can reveal a student's thought processes and conceptual understanding in ways that static text cannot \cite{graesser2005scaffolding,vanlehn2007}. Interactive systems that generate questions can provide insight into a student's critical thinking and problem-solving strategies, thereby capturing evidence of achievement that traditional AES or detectors miss \cite{heilman2010good}. Recent advances in conversational AI and question-generation technology have made this a vibrant area of research for enhancing assessment validity \cite{labadze2023role,burstein2024assessment}.

Our framework builds on this interactive approach. By integrating follow-up questions with AES, we aim to combine the efficiency of automated scoring with the validity of interactive verification, compensating for the instability of detection tools and the scalability challenges of multimodal assessment.

\section{\uppercase{System Design}}

\subsection{System Design Principles}

Our two-stage AI-assisted assessment framework is designed as an integrated system, guiding the process from rubric generation to final feedback. The design is grounded in the following principles.

\begin{enumerate}
  \item \textbf{Consistency based on structured rubrics:} To ensure consistent evaluation and address the low accuracy associated with unconstrained zero-shot grading \cite{piscitelli2025large}, the system uses a structured rubric schema. The LLM is constrained to output scores, justifications, and follow-up questions in a standardized JSON format for each criterion.
  \item \textbf{Transparency and explainability:} Students can review the assignment and rubric in advance and track the entire process, from the initial grade and follow-up questions to the final feedback. Instructors can verify and save the records at each stage to transparently present the evaluation basis.
  \item \textbf{Human in the loop:} Instructors approve and modify AI outputs—including rubrics, scores, questions, and final feedback—throughout the process. This human oversight is crucial for minimizing model errors and incorporating expert judgment.
  \item \textbf{Evidence-based reassessment:} Student responses to follow-up questions serve as additional evidence. This allows the assessment to capture a deeper level of understanding that may not be evident in the initial submission, thereby enhancing both validity and fairness.
\end{enumerate}

\subsection{System Overview}

\begin{figure}[h]
  \centering
  \includegraphics[width=0.7\columnwidth]{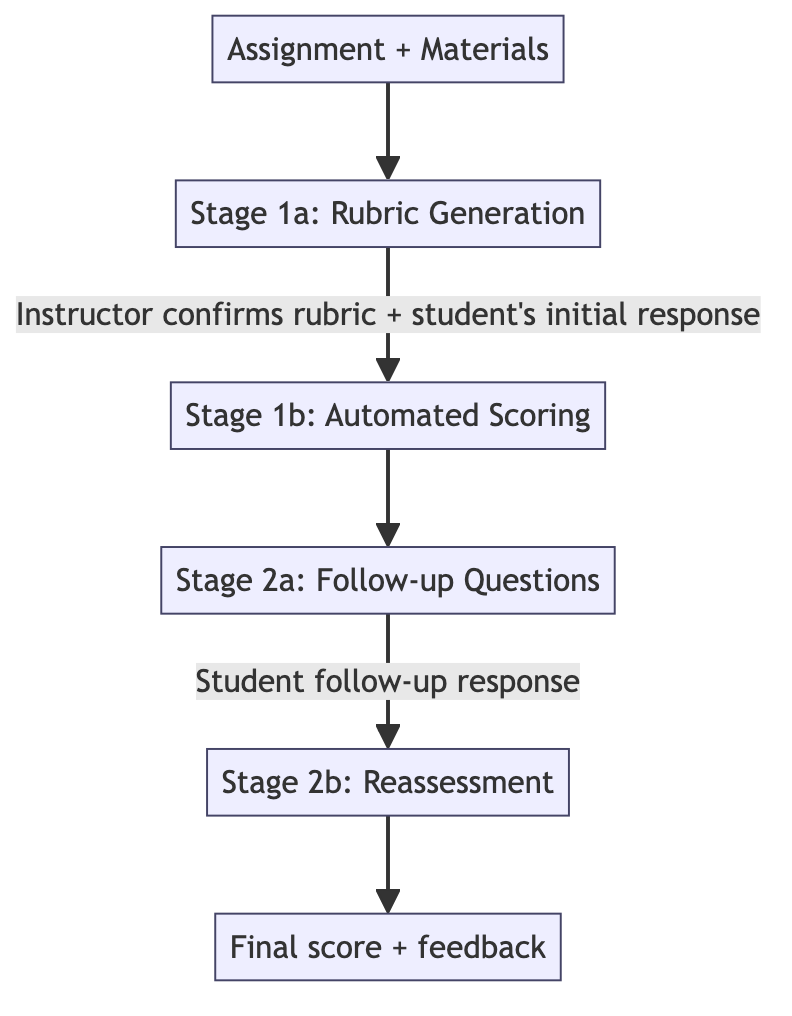}
  \caption{Four-stage system overview: Rubric generation $\rightarrow$ Auto-scoring (initial response) $\rightarrow$ Follow-up questions $\rightarrow$ Reassessment (follow-up response).}
  \label{fig:four_stages}
\end{figure}

The system architecture follows the four-stage process depicted in Figure~\ref{fig:four_stages}. An instructor approval point follows each automated step, ensuring human oversight. We use the Gemini 2.5 Flash model and enforce structured JSON outputs to ensure schema validity for downstream processing. For this pilot, curricular materials were adapted from the Read.Inquire.Write project on Athenian democracy \cite{ReadInquireWrite2024Athens}, and all "student answers" were instructor-authored exemplars.

\definecolor{StronglyDisagree}{HTML}{D53E4F} 
\definecolor{Disagree}{HTML}{F46D43} 
\definecolor{Neutral}{HTML}{E0E0E0} 
\definecolor{Agree}{HTML}{74ADD1} 
\definecolor{StronglyAgree}{HTML}{4575B4} 

\newcommand{\divergentbar}[3]{%
  \color{Neutral}\rule{0.25\dimexpr#1mm\relax}{10pt}%
  \color{Agree}\rule{0.25\dimexpr#2mm\relax}{10pt}%
  \color{StronglyAgree}\rule{0.25\dimexpr#3mm\relax}{10pt}%
}

\subsection{Stage 1a: Automatic Rubric Generation}

\begin{figure}[htbp]
  \centering
  \setlength{\tabcolsep}{3pt}
  \noindent
  \setlength{\FrameRule}{0.4pt}
  \setlength{\FrameSep}{6pt}
  \begin{framed}
    {\footnotesize\ttfamily
      You are an academic expert in assessment design. Using the provided inputs:

      \begin{itemize}[nosep, leftmargin=1.2em, label=\texttt{-}, labelsep=0.4em]
        \item \texttt{course material} (e.g., unit readings or lesson page)
        \item \texttt{syllabus.pdf} (learning objectives, outcomes)
        \item \texttt{assignment prompt} (task description, constraints)
      \end{itemize}

      Generate an analytical rubric with 1--5 performance levels per criterion and weights that sum to 100. Ensure criterion names are concise, operational definitions are specific, and levels are observable.}
  \end{framed}
  \centering\footnotesize (a) Input prompt (abstracted)\par

  \noindent
  \begin{framed}
    \begin{minipage}{\columnwidth}
      \footnotesize\ttfamily
      criteria:\\
      - criterion\_name: "Evidence Use", weight: 25\\
      \hspace*{1em}description: "Select and present appropriate evidence from course sources (1--5), cite accurately, and integrate to support claim."\\
      \hspace*{1em}levels: - level: 5, desc: "Selects most relevant evidence; cites accurately; integrates seamlessly.
      \\[0.5em]
      - criterion\_name: "Claim", weight: 20\\
      \hspace*{1em}description: "Present a clear and logical claim that answers the task prompt."\\
      \hspace*{1em}levels: - level: 3, desc: "Clear claim, but precision/persuasiveness limited; critique somewhat shallow.
      \\[0.5em]
      - criterion\_name: "Historical Context", weight: 15\\
      \hspace*{1em}description: "Reflect complexity and limits of Athenian democracy (e.g., restricted citizenship, inequality)."\\
      \hspace*{1em}levels: - level: 1, desc: "Little to no understanding; oversimplifies or includes factual errors."
    \end{minipage}
  \end{framed}
  \centering\footnotesize (b) Generated output object example (subset; translated; illustrative)\par

  \caption{Stage 1a inputs and outputs. (a) Input prompt (abstracted). (b) Generated output object example (subset; translated; illustrative).}
  \label{fig:stage1a_quad}
\end{figure}

Figure~\ref{fig:stage1a_quad} illustrates automatic rubric generation from assignment related inputs: the analytic rubric consists of criteria with weights (summing to 100) and description and level descriptors for each level 1 to 5. These descriptors serve as anchors for targeted questioning and evidence-based reassessment in later stages. This automatic generation alleviates instructor burden while preserving transparency and consistency across students.

\subsection{Stage 1b: Rubric-Based Automatic Scoring}

After a student submits their assignment, the system applies the instructor-approved rubric to perform an initial automated scoring. For each criterion, the model generates a score (1-5) and a textual justification for that score. The instructor reviews these outputs, adjusting scores or rationales as needed to ensure accuracy and fairness before the results are used to generate follow-up questions.

\subsection{Stage 2a: Follow-up Question Drafting}

Using the initial scores and justifications, the system drafts up to three customized follow-up questions designed to elicit further evidence of the student's understanding. Two types of prompts are used: (1) \textbf{Evaluative supplementary questions} to resolve ambiguous scores by requesting examples, elaborations, or citations tied to specific criteria; and (2) \textbf{Authenticity verification questions} to probe transfer, rephrasing, or contextual adaptation, checking for consistency and explainability. The instructor reviews and can edit these questions for clarity, difficulty, and pedagogical alignment before they are sent to the student.

\subsection{Stage 2b: Reassessment Reflecting Student Responses}

\begin{figure}[htbp]
  \centering
  \begin{framed}
    {\footnotesize\ttfamily
      You are an instructor-facing assessment reviewer. Using the provided inputs:

      \begin{itemize}[nosep, leftmargin=1.2em, label=\texttt{-}, labelsep=0.4em]
        \item \texttt{course material} (e.g., unit readings or lesson page)
        \item \texttt{assignment submission} (assignment rubric, original submission, auto-generated scores)
        \item \texttt{follow-up questions and responses}
      \end{itemize}

      Generate a final feedback, a final score, and justifications for each criterion in a structured format.}
  \end{framed}
  \centering\footnotesize (a) Stage 2b reassessment prompt (abstracted)\par

  {\footnotesize
    \begin{tabular*}{\columnwidth}{@{\extracolsep{\fill}}p{0.3\columnwidth}p{0.6\columnwidth}}
      \toprule
      \textbf{Criterion} / \textbf{Score} & \textbf{Supporting rationale} \\
      \midrule
      Evidence Use \\ ($3\rightarrow4$) & The student worked to support the claim with course sources (1, 2, 3). The follow-up response demonstrated a precise understanding of sources 4 and 5 and laid out concrete ways they reinforce the argument, revealing substantial growth in evidence use and justifying the higher score. \\
      \midrule
      Reasoning \& Analysis \\ ($4\rightarrow5$) & The student clearly explained how sources 1, 2, and 3 substantiate the claim, notably analyzing the limits of critique within Athenian democracy through source 2 and relating the Delian League to democratic practice. The follow-up leveraged source 4 to illustrate how \"customary and traditional viewpoints\" constrained democracy, deepening the analysis and providing the decisive basis for the score increase. \\
      \bottomrule
    \end{tabular*}}
  \centering\footnotesize (b) Example score adjustments and rationale (subset; illustrative)\par

  \caption{Stage 2b reassessment artifacts derived from the pilot prototype. Panel (a) abstracts the system prompt guiding final scoring. Panel (b) shows a subset of score adjustments and the corresponding rationale communicated to the learner.}
  \label{fig:stage2b_prompt}
\end{figure}

Once the student provides responses to the follow-up questions, the system initiates a reassessment. This iterative process aligns with findings by \cite{bhojan2025reflexai}, who showed that repeated queries and \say{Chain-of-Thought} interaction significantly improve feedback consistency and student reflection. As shown in Figure~\ref{fig:stage2b_prompt}, the model is provided with the complete context: the original submission, the rubric, the initial scores and justifications, and the follow-up question–response pairs. It is then prompted to recalculate the score for each criterion based on the totality of the evidence and to generate new justifications for any changes. For instance, as shown in Panel (b), a score for \say{Evidence Use} might increase from 3$\rightarrow$4 if a student's follow-up response demonstrates a more precise understanding of how specific sources support their claim. The system also drafts comprehensive final feedback for the student, summarizing the initial assessment, how the follow-up responses provided new evidence, and suggestions for improvement. All reassessment outputs are subject to final instructor approval.

\section{\uppercase{Pilot Study Design}}

We conducted an instructor-centered pilot study to evaluate the feasibility of the proposed two-stage assessment framework and identify areas for improvement. The study focused exclusively on instructor perceptions and did not involve student data or participation.

\begin{figure}[htbp]
  \centering
  \setlength{\tabcolsep}{3pt}
  \begin{tabular}{@{}m{0.48\columnwidth} m{0.48\columnwidth}@{}}
    {\centering\includegraphics[width=\linewidth,keepaspectratio]{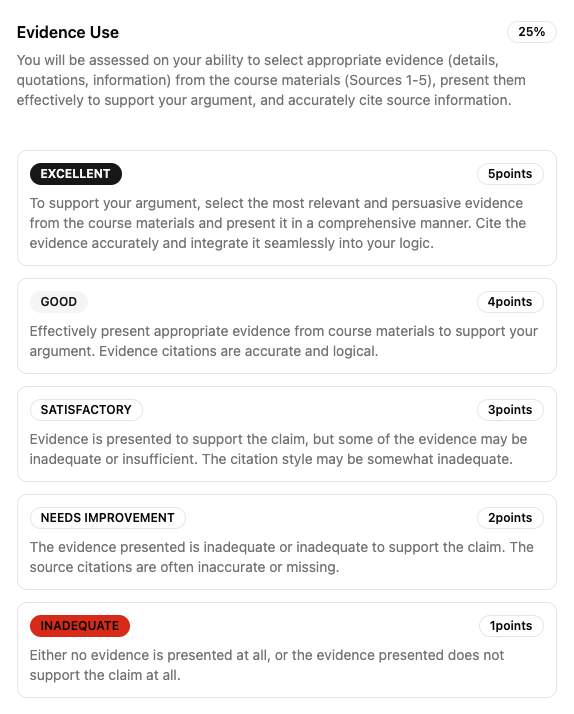}\par} &
    {\centering\includegraphics[width=\linewidth,keepaspectratio]{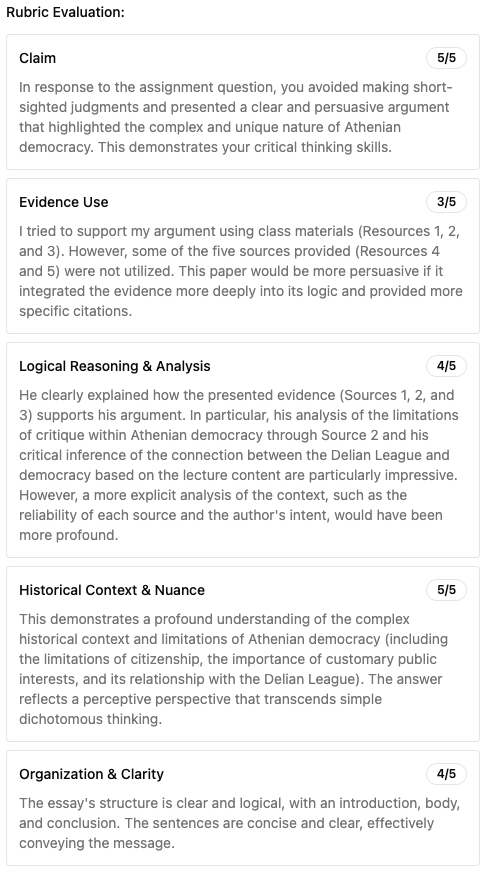}\par}                                                                              \\
    {\centering\footnotesize (1) Stage 1a: Rubric Generation\par}                                                               & {\centering\footnotesize (2) Stage 1b: Rubric-Based Automatic Scoring\par} \\
    {\centering\includegraphics[width=\linewidth,keepaspectratio]{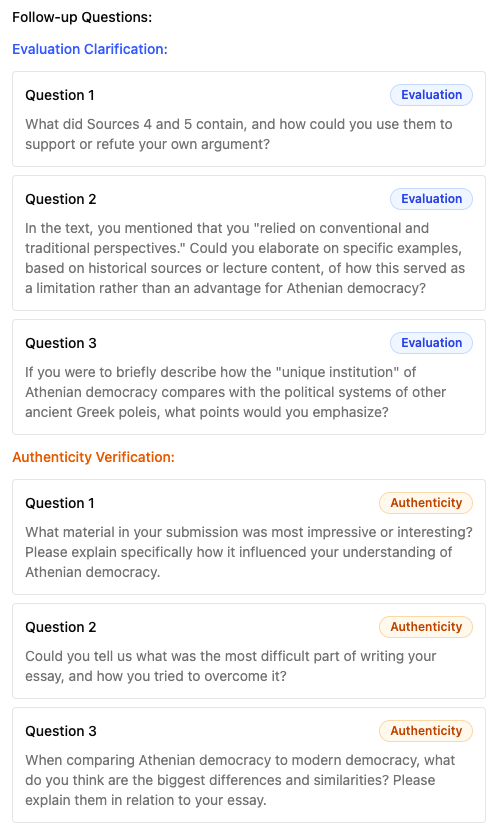}\par} &
    {\centering\includegraphics[width=\linewidth,keepaspectratio]{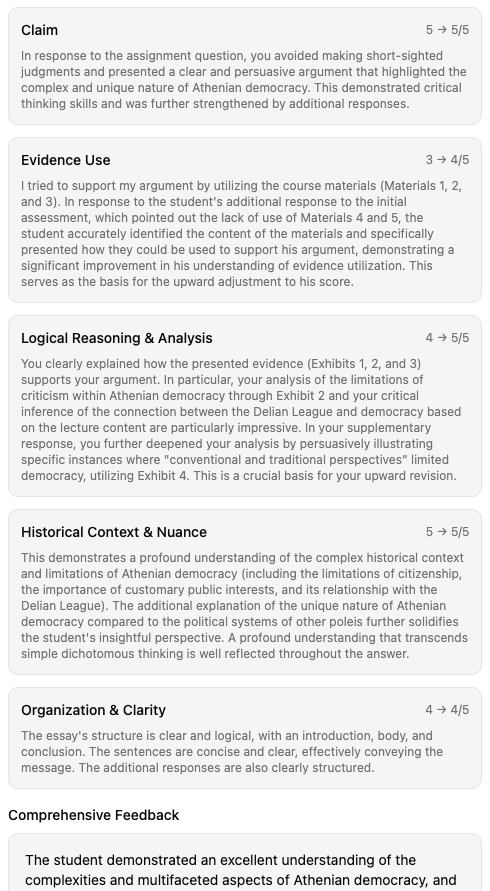}\par}                                                                              \\
    {\centering\footnotesize (3) Stage 2a: Follow-up Questions\par}                                                             & {\centering\footnotesize (4) Stage 2b: Reassessment\par}                   \\
  \end{tabular}
  \caption{Pilot study procedure screenshots\protect\footnotemark.}
  \label{fig:pilot_procedure}
\end{figure}
\footnotetext{Screenshots translated using Google Translate on the website.}

\subsection{Participants}

Nine university instructors from various academic disciplines participated in the study. Although the sample size was relatively small, prior research suggests that usability evaluations with approximately five participants can be sufficient to identify the majority of usability problems \cite{nielsen1993mathematical}. Each instructor tested the system using a narrative assignment from one of their courses. In place of real student work, they used instructor-authored exemplar responses to evaluate the system's outputs. The diverse backgrounds of the participants were intended to provide a broad perspective on the applicability of the framework, reflecting the widespread need among educators for new assessment methods in the age of LLMs \cite{morjaria2023examining}.

\subsection{Procedure}

The pilot study followed a six-step procedure, with UI screenshots illustrated in Figure~\ref{fig:pilot_procedure}:

\begin{enumerate}
  \item \textbf{Assignment Upload:} Each instructor uploaded an assignment and relevant course materials into the system.
  \item \textbf{Automatic Rubric Generation (Stage 1a):} The system generated an analytic rubric based on the provided materials.
  \item \textbf{Rubric-Based Automatic Scoring (Stage 1b):} The system scored the instructor-authored exemplar submissions against the rubric, allowing for an evaluation of its reliability and validity in line with recent AES research \cite{yavuz2025utilizing}.
  \item \textbf{Stage 2a: Follow-up Question Drafting.} Based on the initial scoring, the system drafted follow-up questions designed to verify understanding, a key strategy for ensuring authenticity \cite{fenton2024reconsidering}.
  \item \textbf{Stage 2b: Reassessment.} The system performed a reassessment using instructor-authored proxy responses to the follow-up questions, simulating how new evidence would update scores and feedback.
  \item \textbf{Instructor Review and Survey:} Instructors reviewed all system-generated outputs (rubric, scores, questions) and completed a survey about their experience.
\end{enumerate}

This procedure was designed to gather instructor feedback on the usability, fairness, and validity of the system.

\subsection{Data Collection}

\begin{table*}[htbp]
  \caption{Instructor survey areas and key items.}
  \label{tab:survey}
  \begin{tabular*}{\textwidth}{@{}p{0.4\textwidth} @{\extracolsep{\fill}} p{0.55\textwidth}@{}}
    \toprule
    \textbf{Key Evaluation Item} & \textbf{Detailed Question Example} \\
    \midrule
    \multicolumn{2}{@{}l}{\textbf{System Usage}} \\
    Usability & \say{Was the system generally easy to use?} \\
    Time required & \say{Was the time required to use the system appropriate?} \\
    \addlinespace
    \multicolumn{2}{@{}l}{\textbf{Stage 1a: Rubric Generation}} \\
    Fairness: Applicability of the same criteria to all students & \say{Does the automatically generated rubric apply the same criteria to all students?} \\
    Content Validity: Reflection of course objectives & \say{Does the rubric align with the course's educational objectives and learning content?} \\
    \addlinespace
    \multicolumn{2}{@{}l}{\textbf{Stage 1b: Rubric-Based Automatic Scoring}} \\
    Fairness: Consistent scoring for similar answers & \say{Does the automatic scoring assign consistent scores to similar answers?} \\
    Construct Validity: Reflection of student's actual understanding & \say{Do the automatic scoring results reflect the student's actual understanding?} \\
    \addlinespace
    \multicolumn{2}{@{}l}{\textbf{Stage 2a/2b: Follow-up Questions and Reassessment}} \\
    Fairness: Consistent and transparent criteria & \say{Are the follow-up questions and reassessment provided consistently and transparently?} \\
    Content Validity: Relevance to learning objectives & \say{Are the follow-up questions and reassessment highly relevant to the learning objectives?} \\
    Construct Validity: Revealing the learner's reasoning process & \say{Do the follow-up questions and reassessment help reveal the student's actual understanding and reasoning?} \\
    \addlinespace
    \multicolumn{2}{@{}l}{\textbf{Overall Evaluation}} \\
    System's contribution to fairness and validity & \say{Do you believe the system contributes to ensuring evaluation fairness and validity?}\\
    \bottomrule
  \end{tabular*}
\end{table*}

We collected data from two primary sources: the AI-generated artifacts from the system (e.g., rubrics, scores, questions) and the survey responses from the nine instructors. The survey, detailed in Table~\ref{tab:survey}, was structured into four sections to assess perceptions of fairness, content validity, and construct validity at each stage of the process \cite{weber2023testing}.

\subsection{Significance}

Although this pilot was conducted on a small scale without student data, it provides significant early-stage insights. The study allowed us to gather expert feedback on the perceived fairness of the rubric generation and automated scoring, assess the validity of the follow-up questions, and identify key directions for system improvement. This demonstrates a new possibility for addressing the crisis of fairness and validity in educational assessment in the era of LLMs \cite{morjaria2023examining,yavuz2025utilizing}.

\section{\uppercase{Results}}

We analyzed survey responses from all nine participating instructors (N=9)\footnote{The parenthetical notations (P1–P9) in quotations are anonymized participant identifiers.}. All items were rated on a 5-point Likert scale (1=Strongly disagree, 5=Strongly agree). The quantitative results and illustrative qualitative feedback are summarized below.

\begin{table*}[htbp]
  \caption{Instructor survey results (N=9): Likert means, positive-agreement rates, and response distributions.}
  \label{tab:results}
  \begin{tabular*}{\textwidth}{@{\extracolsep{\fill}}llccc@{}}
    \toprule
    \textbf{Area} & \textbf{Evaluation Item} & \textbf{Response Distribution} & \textbf{Mean} & \textbf{Agree \%} \\
    \midrule
    \textbf{System Usage} & Usability & \divergentbar{0}{66.7}{33.3} & 4.3 & 100.0\% \\
    & Time Appropriateness & \divergentbar{11.1}{44.4}{44.4} & 4.3 & 88.9\% \\
    \addlinespace
    \textbf{Rubric Generation} & Fairness & \divergentbar{11.1}{44.4}{44.4} & 4.3 & 88.9\% \\
    & Content Validity & \divergentbar{0}{55.6}{44.4} & 4.4 & 100\% \\
    \addlinespace
    \textbf{Auto-Scoring} & Fairness & \divergentbar{11.1}{44.4}{44.4} & 4.3 & 88.9\% \\
    & Construct Validity & \divergentbar{44.4}{44.4}{11.1} & 3.7 & 55.6\% \\
    \addlinespace
    \textbf{Follow-up Questions} & Fairness & \divergentbar{0}{44.4}{55.6} & 4.6 & 100\% \\
    & Content Validity & \divergentbar{0}{33.3}{66.7} & 4.7 & 100\% \\
    & Construct Validity & \divergentbar{22.2}{66.7}{11.1} & 3.9 & 77.8\% \\
    \addlinespace
    \textbf{Overall} & Contribution to Fairness/Validity & \divergentbar{0}{44.4}{55.6} & 4.6 & 100\% \\
    \bottomrule
  \end{tabular*}
  \begin{center}
    \begin{tabular}{@{}c@{}}
      \color{Neutral}\rule{10pt}{10pt} Neutral \quad
      \color{Agree}\rule{10pt}{10pt} Agree \quad
      \color{StronglyAgree}\rule{10pt}{10pt} Strongly Agree
    \end{tabular}
  \end{center}
\end{table*}

As shown in Table~\ref{tab:results}, instructors rated the system's overall \textbf{usability} (M=4.3) and \textbf{time appropriateness} (M=4.3) favorably, with positive agreement rates of 100\% and 88.9\%, respectively. This suggests the framework has an acceptable learning curve and operational overhead. One instructor noted, \say{The UI is intuitive and conveniently organized} (P2), though another commented on performance, stating, \say{it was slow overall} (P3).

\textbf{Rubric generation} (Stage 1a) was perceived as strong in both \textbf{fairness} (M=4.3, 88.9\% agreement) and \textbf{content validity} (M=4.4, 100\% agreement). Participants appreciated that \say{the same criteria were applied to all students} (P3), which supports procedural fairness. However, they also observed that rubric quality was dependent on the input materials, with one noting, \say{the more detailed the syllabus, the more detailed the rubric} (P4).

For \textbf{automated scoring} (Stage 1b), instructors found the system to be \textbf{fair} in its consistency (M=4.3, 88.9\% agreement), agreeing that it assigned \say{consistent scores for similar answers} (P3). However, its \textbf{construct validity} received the lowest rating of any metric (M=3.7, 55.6\% agreement). One participant captured this sentiment, citing \say{limitations in reflecting the student's level of understanding in detail} (P3). This result highlights a key tension: while consistent, the automated scoring was seen as insufficient for capturing the depth of student reasoning on its own.

The \textbf{follow-up questions and reassessment} (Stage 2) received the highest ratings, particularly for \textbf{content validity} (M=4.7) and \textbf{fairness} (M=4.6), with 100\% positive agreement for both. Instructors confirmed the questions \say{were consistent with the learning objectives} (P4) and found the process \say{transparent} because \say{the criteria are subdivided} (P5). \textbf{Construct validity} was also rated positively (M=3.9, 77.8\% agreement), though instructors suggested a key area for improvement. One participant recommended that \say{the difficulty of the questions needs to be differentiated according to the student's level} (P3), pointing toward the need for adaptive questioning.

\textbf{Overall}, instructors unanimously agreed (M=4.6, 100\% agreement) that the two-stage framework contributes positively to assessment fairness and validity. One participant summarized the sentiment well, stating that while \say{it's hard to see it as decisive... it's definitely helpful} (P4). These exploratory findings suggest the framework's strengths lie in its transparent, consistent scoring and goal-aligned questioning, while also identifying the construct validity of initial auto-scoring and the need for adaptive difficulty as key priorities for future development.

\section{\uppercase{Discussion}}

This instructor pilot study explored a two-stage assessment framework designed to address the challenges of fairness and validity in the LLM era. The findings, while exploratory, reveal a crucial tension between procedural fairness and construct validity, offering valuable insights for the future design of AI-assisted assessment tools.

Instructors valued the automated rubric generation (Stage 1a) for its ability to enforce procedural fairness through the consistent application of transparent criteria. This aligns with the principle that fairness requires an equitable opportunity for all students to demonstrate their learning \cite{glaser1997assessment}. However, the feedback also underscored that the system's effectiveness is contingent on well-structured input from educators, highlighting the symbiotic relationship between course design principals and AI tools.

The automated scoring (Stage 1b) further illuminated the fairness-validity tension. While instructors perceived the scoring as fair due to its consistency across similar responses, they rated its construct validity lower, noting that it tended to reward surface-level features over deep comprehension. This echoes long-standing critiques of AES systems \cite{deane2013relation,yavuz2025utilizing}. While \cite{todorov2025evaluating} suggest that full fine-tuning of models yields the highest performance for short-answer scoring, this approach is resource-intensive and difficult to scale across diverse assignments. Our framework offers a more efficient alternative: by keeping the human in the loop and using dynamic follow-up questions, we achieve high validity without the need for per-task model training.

The interactive follow-up questions (Stage 2) emerged as a critical component for resolving this tension. By probing for student reasoning, this stage was seen as a powerful tool for strengthening validity. Instructors found the questions to be fair, relevant, and well-aligned with learning objectives, consistent with research on the pedagogical value of inquiry-based interaction \cite{graesser2005scaffolding,heilman2010good}. This interactive element provided a mechanism to gather the "process evidence" that automated scoring missed, thereby moving the assessment closer to a true measure of student knowledge \cite{chai2024grading}. The call for adaptive difficulty in these questions points to a clear direction for future work to further enhance construct validity.

Taken together, our findings suggest that the stages of the framework have a complementary relationship. Rubric generation establishes a foundation of fairness and alignment; automated scoring offers efficiency and consistency; and interactive questioning supplements validity by probing for authentic understanding. The key insight from this pilot is that a blended approach is necessary: purely automated methods may achieve consistency at the cost of depth, while interactive components can restore the balance by gathering more authentic evidence of student learning.

The findings are exploratory, based on a small sample of instructors and did not include actual student data or responses. Therefore, the results should be interpreted as preliminary indicators of feasibility and user perception, not as generalizable evidence of the system's effectiveness. Future work must validate this framework in live classroom settings to analyze its impact on student learning and instructor workload. This will involve implementing adaptive question-generation algorithms and exploring multimodal extensions (e.g., analyzing speech and gaze) to further strengthen the assessment of authenticity and validity.

\section{\uppercase{Conclusions}}
\label{sec:conclusion}

This paper presented a two-stage, AI-assisted evaluation framework designed to mitigate the challenges of fairness and validity in assessing student work in the LLM era. The framework integrates rubric-based automated scoring with targeted follow-up questions, thereby combining the efficiency of automation with the authenticity of interactive verification. Our exploratory pilot study with university instructors suggests that this hybrid approach is a feasible and promising direction for enhancing the integrity of narrative assessments.

The primary contribution of this work is demonstrating how a complementary design can balance the often-competing demands of fairness and validity. The automated scoring stage provides procedural fairness and consistency, while the interactive verification stage strengthens construct validity by eliciting direct evidence of student reasoning. This synthesis operationalizes theoretical perspectives on validity \cite{messick1995validity} and fairness \cite{glaser1997assessment} into a practical, instructor-in-the-loop workflow. This framework offers educators a viable alternative to unreliable AI detection tools and resource-intensive manual methods, promoting both academic integrity and student trust in the assessment process.

The study's limitations, including its small, instructor-only sample, mean the findings are preliminary. The next essential phase of this research is to deploy and evaluate the framework in real classroom environments with student participation. Specifically, we plan to conduct large-scale field studies during the academic semester to track students' learning outcomes and changes in perception over time. Future work will also focus on refining the adaptive difficulty of the question-generation engine and exploring the potential of multimodal data to create an even more robust and authentic assessment experience.

Ultimately, this pilot study indicates that combining the strengths of automation with the irreplaceable value of human-like interaction may provide a pathway toward fairer, more valid, and more meaningful assessment. By continuing to refine and test this framework, we aim to contribute both theoretically and practically to the ongoing rethinking of educational evaluation in a time of rapid technological change.



\bibliographystyle{apalike}
{\small
\bibliography{references}}

\end{document}